\documentclass{iopart}
\usepackage{iopams}

\usepackage{psfig}

\newcommand{\be}{\begin{equation}}
\newcommand{\ee}{\end{equation}}
\newcommand{\bea}{\begin{eqnarray}}
\newcommand{\eea}{\end{eqnarray}}

\newcommand{\bm}{\mathbf}

\begin{document}

\title{Granular gases: dynamics and collective effects}
\author{
A. Barrat\dag, E. Trizac\ddag,  and M.H. Ernst\S}
\address{\dag Laboratoire de Physique Th\'eorique
(UMR 8627 du CNRS), B\^atiment 210,
Universit\'e Paris-Sud, 91405 Orsay, France}
\address{\ddag Laboratoire de Physique Th\'eorique et Mod\`eles Statistiques
(UMR 8626 du CNRS), B\^atiment 100,
Universit\'e Paris-Sud, 91405 Orsay, France}
\address{\S Instituut voor Theoretische Fysica, Universiteit Utrecht,
Postbus 80.195, 3508 TA Utrecht, The Netherlands}

\begin{abstract}
We present a biased review of some of the most "spectacular" 
effects appearing in the dynamics of granular gases where the dissipative 
nature of the collisions leads to a rich phenomenology,
exhibiting striking differences with equilibrium gases. 
Among these differences, the focus here is on the illustrative
examples of  ``Maxwell Demon''-like experiment, modification
of Fourier's law, non-equipartition of energy and non-Gaussianity
of the velocity distributions. The presentation remains as 
non technical as possible.
\end{abstract}
%\pacs{}
\maketitle

\section{Introduction}
\label{sec:intro}

Coulomb, Faraday, Huygens or Reynolds are among the prominent founding fathers 
of the study of granular materials.  Subsequently this field has mostly 
been investigated by engineers.  During the last 15 years however, 
physicists seem to have rediscovered the field and gradually became 
more involved again, partly due to the increase of computer 
resources \cite{book1,book2}.
Beyond fashion effects and important industrial stakes the behavior 
of granular matter in its own right is remarkably rich, and often 
resists understanding \cite{JNB}.

The corresponding systems are nevertheless simple to define. Broadly speaking,
they are composed of macroscopic compounds (with sizes larger than a fraction
of millimeter). Two crucial properties of granular matter directly follow from
this size constraint: first, ordinary temperature is irrelevant, as may be
appreciated  by comparing the typical gravitational energy of a grain to the
thermal agitation energy.  Second, the interactions between grains are
dissipative. They involve complex macroscopic processes such as fracture,
friction, internal vibrations etc which contribute to dissipate kinetic
energy.  The mechanisms for exploring configuration space are therefore
unusual and any dynamics results from an external drive. The resulting
stationary states often differ from those observed in conservative
systems (such as molecular gases, ordinary liquids, or colloidal suspensions)
with sometimes spectacular or unexpected effects. The purpose of the present
paper is to present in a concise and non technical manner four manifestations
of those differences. We will restrict the discussion to the so-called gaseous
state of granular matter, where a rapid flow is obtained by a violent and
sustained excitation.  This does not necessarily mean that the total density
has to be small.  In this regime (opposed to the quasi-static limit which has
been the object of intense research and where solid or liquid-like behavior may be observed), the only contacts between grains occur during
collisions. The corresponding "gases" have been studied experimentally, but
also with numerical and analytical tools, and provide an ideal
contact and ground for  comparison between the various
approaches \cite{book1,book2}.

This paper is organized as follows: section \ref{sec:exp} describes the
general experimental and theoretical framework of granular gases.  Section
\ref{sec:demon}, \ref{sec:Fourier}, \ref{sec:mixtures}, \ref{sec:velocity}
then present four striking consequences of the dissipative nature of 
collisions in granular
gases. For more details, the interested reader is referred to 
comprehensive books, such as Ref \cite{book1,book2}, to reviews
\cite{Kadanoff,Duftyr,Goldh_review} and to the references provided therein.

%%%%%%%%%%%%%%%%%%%%%%%%%%%%%%%%%%%%%%%%%%%%%%%%%%%%%%%%%%%%%%%%%%%%%%
\section{Experiments, models and theoretical approaches}
\label{sec:exp}

\subsection{Experiments}
A granular gas is typically obtained by enclosing sand or balls made of glass,
steel, brass, ceramic beads, etc in a container, which is subsequently vigorously shaken. The energy injected at the boundaries compensates for the dissipative collisions
\cite{rque}, and allows the grains (particles) to follow ballistic
trajectories between collisions. The experimentally challenging aspects deal
with measurements of  different relevant quantities (e.g. local density,
velocity distribution from which the velocity field may be extracted etc).
This requires sophisticated detection devices 
(ultra-fast cameras, diffusive wave spectroscopy, magnetic resonance, 
positron based techniques...) \cite{Rouyer,Wildman2,Losert,Aranson}. The
most frequently employed experimental set-ups are made up either of
cylindrical containers \cite{Wildman2} (which may block direct
visualization), or of thin cages with transparent boundaries\cite{Rouyer}, which
makes the system quasi two-dimensional.  In order to obtain reproducible and
reliable results, much effort has been paid to control the size distribution,
with spherical beads of typical millimetric size.

\subsection{Modeling}
Two features are essential to capture the behavior of granular gases: the
excluded volume on the one hand (hard core effect), and the dissipative nature
of collisions on the other hand. The simplest approach incorporating these two
aspects is the paradigmatic inelastic hard sphere model \cite{Haff,Campbell}.
In this modeling, spherical grains that do not interact at distance larger
than their diameter (with forbidden overlaps), undergo momentum conserving but
dissipative (inelastic)
collisions. These collisions are assumed instantaneous so 
that events involving more than two partners may be neglected. 
Suppose we have a binary mixture, where $i$ labels both 
the particle and the species in the mixture, and where 
$m_i = \{m_1,m_2\}$. For two grains $ i$ and $j$ belonging to 
species $i$ and $j$ respectively, the post-collision velocities 
(denoted with primes) read 
\be
%\begin{eqnarray}
\bm{v}_i'= \bm{v}_i - \frac{m_{j}}{m_i+m_j} (1+\alpha_{ij})
(\widehat{\bm{\sigma}}\cdot \bm{v}_{ij})\widehat{\bm{\sigma}} 
\label{eq:colla}
%\bm{v}_2'= \bm{v}_2 + \frac{m_{1}}{m_1+m_2} (1+\alpha_{12})
%(\widehat{\bm{\sigma}}\cdot \bm{v}_{12})\widehat{\bm{\sigma}},
%\label{eq:collb}
%\end{eqnarray}
\ee
where $i$ and $j$ taking the values 1 or 2.
Here $\widehat{\bm\sigma}$ is the center-to-center unit vector (oriented
$i\to j$ or $j\to i$) and $\bm{v}_{ij}=\bm{v}_{i}-\bm{v}_{j}$ is the
pre-collision value of the relative velocity. In 
Eq.  (\ref{eq:colla})  $\alpha_{ij}$ is the coefficient of normal restitution
associated with the pair $(ij)$. The collision dissipates kinetic energy for
$\alpha_{ij}<1$, whereas $\alpha_{ij}=1$ corresponds to the elastic
(conservative) case. For the sake of simplicity, the latter coefficient is
taken independent of $\bm{v}_{ij}$ or of the impact parameter of the
collisions. While this is certainly  an oversimplification of the experimental
reality, it appears that thorough measurements of the restitution 
coefficient are
scarce under conditions relevant for the study of bulk properties
\cite{Foerster,Labous,Louge,King}. Bearing these limitations in mind, the
inelastic hard sphere model with a constant coefficient of normal restitution is
useful as a minimal approach to understand the rich phenomenology of granular
gases, since it not only facilitates numerical investigations but also allows
for analytical studies. In particular, the next sections will show how the
striking effects of inelasticity are captured by this simple model.

More realistic models have also been proposed, where the particles 
have rotational degrees of freedom, and where there is in addition 
to normal restitution also tangential restitution \cite{LudingBlumen},
with sticking and sliding friction, restitution 
coefficients depending on the velocities or visco-elastic interactions
\cite{book1,book2,Foerster,Walton,Brilliantov,Luding_pre,Falcon}.

\subsection{Analytical approaches}

>From the theoretical point of view, the methods range from a microscopic
description using kinetic theory, to continuum mechanics-like approaches
(hydrodynamics), aiming at establishing the evolution equations governing the
dynamics of suitably defined coarse-grained fields (density, momentum, kinetic
energy, etc...)~\cite{Duftyr}.  In a molecular gas or an ordinary liquid, the
validity of the hydrodynamic approach relies on the existence of conserved
quantities (collisional invariants) among which the kinetic energy plays a key
role. In a granular gas, this approach seems {\it a priori} questionable due
to dissipation. In addition, new length and time scales emerge, that may
equally well interfere with the microscopic and macroscopic scales
\cite{controversy}.

%%%%%%%%%%%%%%%%%%%%%%%%%%%%%%%%%%%%%%%%%%%%%%%%%%%%%%%%%%%%%%%%%%%%%%
\section{From the Maxwell Demon...}
\label{sec:demon}

In a celebrated thought experiment, James Clerk Maxwell described in 1871 a
demon capable of separating slow from fast molecules in a gas, in order to
create a "hot" compartment and a "cold" one.  Many physicists -- among which
L. Brillouin -- have contributed to exorcize this demon: in an equilibrium gas,
such a spontaneous separation is impossible. However, a granular gas is not an
equilibrium system, but is driven by a continuous supply of energy. Several
groups have shown that under those circumstances, a spontaneous separation
reminiscent of that put forward by Maxwell could be realized \cite{demon}.

The required experimental set-up is simple: the confining box is divided 
in two identical compartments that may communicate through a hole.  
The box is then filled with particles of a granular material
and brought in a gas-like state by vertical shaking. For strong
shaking the grains fill the two halves of the box symmetrically, but
upon decreasing the agitation, a critical threshold is met 
below which the above symmetry is broken. One compartment becomes
more populated than the other; the grains suffer more collisions there
and dissipate more kinetic energy. Identifying the mean kinetic energy
and the temperature by analogy with  the terminology of 
molecular gases (see section \ref{sec:Fourier}),
one then obtains a rather dense and "cold" compartment, 
coexisting with a more dilute and "hot" counterpart. 
This phenomenon explains the name "Maxwell Demon", often used to 
refer to the previous situation \cite{Eggers,rque2}. The 
second law of thermodynamics is nevertheless
not violated ! In contradistinction with the molecules Maxwell
had in mind, the grains are here macroscopic and may absorb and dissipate
energy.

The left/right asymmetry may be anticipated on simple grounds. Due to the
inelasticity of collisions, a dense region in which more collisions occur will
see its mean kinetic energy decrease. If such a fluctuation leads to an
increase of density in one of the compartments, the grains will subsequently
escape at a lower rate. Conversely, in the other compartment, the mean energy
will increase which facilitates the escape. The fluctuation is amplified and
may overcome the energy input from the base if the shaking is not strong
enough, leading to a breakdown of symmetry. Translating the above heuristic
argument into a more quantitative theory is however not an easy task.
Phenomenological approaches have been proposed to refine the argument, with
simplifying assumptions \cite{Eggers,Mikkelsen,Droz}. Equating left-right and
right-left fluxes of grains leads to qualitative agreement with the
experimental observation. The situation where the hole is of large size (e.g.
when the aperture between the two compartments typically extends over half the
box height \cite{Brey_Demon,Molphys}) seems more amenable to analytic
treatment. When the external forcing is sufficiently strong to allow  the
neglect of the gravitation force, a hydrodynamic-like description may be put
forward, leading to excellent agreement with molecular dynamics simulations
of inelastic hard spheres \cite{Brey_Demon,Brey_hydro}. At constant forcing,
the symmetric non equilibrium steady state becomes unstable when the number of
grains exceeds a critical threshold.  One of the compartments then becomes
"colder" and denser than the other one \cite{rque3}.

%%%%%%%%%%%%%%%%%%%%%%%%%%%%%%%%%%%%%%%%%%%%%%%%%%%%%%%%%%%%%%%%%%%%%%
\section{...to Fourier's law}
\label{sec:Fourier}

The hydrodynamic approach -- based on a Chapman-Enskog expansion procedure
starting from the relevant Boltzmann equation 
(see e.g. \cite{Brey_hydro,Sela,Brilli}) --, 
specifies the form of the constitutive relations
between fluxes and gradients. A striking result derived along 
these lines is that Fourier's law (relating the heat flux 
$\bf q$ to the temperature gradient), is modified with respect to 
conservative systems. A new term proportional to the
density gradient must be added to obtain the heat flux and one has
\begin{equation}
{\bm q} = -\kappa \,{\bm \nabla} T - \mu \,{\bm \nabla} n,
\label{eq:mu}
\end{equation}
where $n$ denotes the local density of grains.
The new transport coefficient $\mu$ is positive whereas it vanishes
in a conservative system, as required by the second law of thermodynamics to guarantee that heat flows from hot to cold. On the other hand, the collisions
induce a "heating" of the internal degrees of freedom of the
grains, which is neglected in modeling of granular systems. So,
there is no reason to expect a positive entropy production.

In Eq.(\ref{eq:mu}), the so-called "granular temperature" $T$ has no
thermodynamic foundation, but only a kinetic status. This quantity has no
relation with the usual temperature (irrelevant at the grain scale as
emphasized in the introduction), but is defined as the variance of the
velocity distribution at a given point. The quantity $nT$ is therefore the
local kinetic energy density in the local center-of-mass frame.  This
definition allows for a direct measurement in an experimental system.
Furthermore it coincides with the thermodynamic definition for a system in
equilibrium. The coefficient $\kappa$ is therefore the counterpart of the
thermal conductivity whereas $\mu$ has no analog in a molecular system and is
intrinsically related to the dissipative nature of collisions. The latter
quantity has profound consequences on the behavior of the system, among which
a possible inversion of the granular temperature profile
\cite{Helal,MariaB,Soto,Maria}, a phenomenon that has been observed
experimentally \cite{Clement,Wildman,Blair} and may be understood from a
hydrodynamic argument \cite{MariaB,Soto}.  Consider a granular gas driven by
an oscillating piston located (on average) at height $z=0$.  The energy flux
is clearly directed toward positive $z$ since no energy comes from the empty
region at large height.  Starting from the base where energy is injected, and
increasing $z$, the temperature first decreases since the energy is dissipated
in the bulk. The density of grains may conversely increase or decrease
depending on parameters (gravity and inelasticity).  However, the temperature
$T$ is not a monotonically decreasing function of height but passes through a
minimum before increasing.  This behavior is at first sight inconsistent with
the dissipative nature of collisions, and indicates that heat flows from
``cold'' to ``hot'' ! It is a direct consequence of the fact that $\mu\neq 0$
in a granular system: in the region where $dT/dz>0$, the density decays very
rapidly, which constitutes the dominant contribution to $\bm q$ [see Eq.
(\ref{eq:mu})]. The resulting heat flux has therefore a positive projection
onto the $z$-axis, as it should. Note also that this constraint implies that
the density is either a decreasing function of $z$ \cite{Soto} or may reach a
maximum at a smaller altitude than that where $T$ is minimum \cite{Maria}.

%%%%%%%%%%%%%%%%%%%%%%%%%%%%%%%%%%%%%%%%%%%%%%%%%%%%%%%%%%%%%%%%%%%%%%
\section{Velocity distribution}
\label{sec:velocity}

Another important characteristics of {\it molecular gases} lies in 
the velocity distribution of the molecules. There  collisions 
do not dissipate energy, and the distribution is a Gaussian. 
One can naturally expect this property to break
down for granular gases with dissipative collisions. The
first experimental measurements however were not precise enough 
to show deviations
from a Gaussian. It was only recently that experimental techniques 
became available to determine distributions with pronounced differences from 
Maxwell-Boltzmann
statistics, especially in the high-velocity tails of the distributions
\cite{Olafsen,Losert,Rouyer,Kudrolli,Aranson}. Several authors reported a
stretched exponential law [on the whole range of velocities available, which
covers an accuracy of $4$ to $5$ orders of magnitude for $P(v)$]
\begin{equation}
P(v) \propto \exp[ -(v/v_0)^\nu ] \ ,
\label{eq:pv}
\end{equation}
with various exponents (here $v_0$ is the ``thermal'' r.m.s.  velocity). 
In particular an exponent $\nu$ close to $3/2$ was found
in various experiments \cite{Losert,Rouyer,Aranson}.
This behavior was observed for the
horizontal velocity components of a vertically vibrated 2D system of steel
beads in a wide range of driving frequencies and densities \cite{Rouyer}, but
also in a three dimensional electrostatically driven granular gas
\cite{Aranson}. A question that naturally arises concerns the possible
universality of this distribution.

On a theoretical level, the delicate point concerns the supply of energy to the inelastic hard spheres system, which  compensates for the dissipation caused by inelastic collisions.  The description of driven experiments
indeed requires a forcing mechanism allowing the system to 
reach a steady state.  This task is difficult, but the heating
process and resulting fluidization described by a ``stochastic thermostat''
\cite{Williams,Puglisi,vanNoije,Pre,Montanero,Moon1,Garzo2} has attracted
attention, in particular because it has been shown analytically that $P(v)$
exhibits a high energy tail of the form of Eq. (\ref{eq:pv}) with $\nu=3/2$,
independent of dimension and restitution coefficient \cite{vanNoije}, in
apparent agreement with the experiments. The above model, where an external spatially homogeneous 
white-noise driving force acts on the particles and thus injects energy
through random ``kicks'' between the collisions, is therefore considered to
provide a relevant theoretical framework to quantify the non Gaussian form of
velocity distributions.

The experimental and theoretical conditions are quite different: in the
experiments, the energy is injected at the boundaries, and the system is not
homogeneous; the theoretical model on the other hand considers 
a homogeneous driving by a
white-noise, acting in the bulk of a homogeneous system. 
It turns out however that the agreement between experiment and theory 
for the exponent $\nu$ is somewhat misleading.
For realistic values of the inelasticity ($\alpha>0.7$) and at
the level of the spatially homogeneous
Boltzmann equation, the predicted high-velocity tails for
$P(v)$, decaying as $\exp(-A v^{3/2})$, are only reached for velocities far
beyond the experimentally accessible ones \cite{EPJE}. 
At "thermal" velocities,
$P(v)$ is in fact close to a Gaussian. Of course this does not correspond to a
failure of kinetic theory, but is most likely caused by a 
too simple model for forcing.

Various groups have subsequently tackled the problem by numerical simulations
of molecular dynamics \cite{Brey,Paolotti,PRE_Alain,vanZon,Moon,Zippelius}:
such simulations of inelastic hard disks in a two-dimensional box
allow to use a reasonably realistic energy injection
through vibrating walls at the boundaries of the box,
and to study the effect of the various
parameters (inelasticities, average density). 
The parameters can be adjusted to obtain velocity
distributions close to their experimental counterparts, and a very 
good precision
can be reached. The simulations lead to the conclusions that the distributions
display generically overpopulated tails with respect to a Gaussian; moreover,
the details of the distributions depend on the various parameters (density,
inelasticity, energy injection...) and even on the part of
the system where the distribution is measured (i.e. it depends slightly
e.g. of the distance from the energy injecting boundaries).

>From the kinetic theory point of view, numerous works have also
been devoted to the understanding of the velocity distributions
emerging from the inelastic Boltzmann equation. Simplified
models such as the Maxwell model may allow for
analytical solutions \cite{balda,brito-ernst,maxwell}, while numerical
resolution is often used in addition to partial solutions
which allow for the prediction e.g. of the high-velocity tails
behavior \cite{Montanero,inprep}. These high-velocity tails
generically display stretched exponential behaviors with an exponent
depending on the details of the model, while power-law velocity
distributions may also be obtained in marginal situations
\cite{balda,brito-ernst,maxwell,inprep}.

The present consensus emerging from these various studies tends to the
conclusion of absence of universality in the velocity distributions: various
experimental conditions, various energy injection modes lead to different
distributions. Of course, these conclusions are based on numerical studies of
simplified models, so that the question of universality can still be
considered as open from an experimental point of view. The only common point
seem to concern the overpopulation of the high-velocity tails with respect to
a Gaussian. To our knowledge, no simple argument however exists to justify
this phenomenon.

%%%%%%%%%%%%%%%%%%%%%%%%%%%%%%%%%%%%%%%%%%%%%%%%%%%%%%%%%%%%%%%%%%%%%%
\section{Breakdown of equipartition in mixtures}
\label{sec:mixtures}

Before concluding, let us turn briefly our attention to mixtures. For a
mixture of molecular gases in equilibrium, all species have the same
temperature, irrespective of their mass, size, or density. One may wonder if
such an equipartition still holds in a granular gas. Two groups have addressed
this question experimentally \cite{Wildman2,Feitosa}, and their results for
binary mixtures clearly demonstrate that equipartition does not hold. The mean
kinetic energy of heavy grains is larger than the one of the light component.
These studies have stimulated numerical and analytical investigations, mostly
centered on the kinetic theory of inelastic hard spheres
\cite{jarek,Pagnani,GranMatt,Paolotti,Dahl,Biben,Talbot}.

In simplified situations of homogeneous systems (with no forcing \cite{Garzo}
or white noise driving \cite{GranMatt}, as well as in shear flows
\cite{Montanero2}) analytical progress is possible. The distinction between
collisions of particles of the same species or of different species leads to
closed equations for the granular temperatures of the two components. Solving
these equations allows then to investigate the dependence of temperature
in-equipartition on the various parameters of the problem (mass ratio, size
ratio, densities, in-elasticities...), which can be varied more easily than in
experiments.  More realistic molecular dynamics simulations with e.g. energy
injection at the boundaries have as well been carried out
\cite{Pagnani,Paolotti,Dahl,Talbot,PRE_Alain}.  The results of these various
investigations are in qualitative agreement with the experiments, despite the
simplifications implied by the modelization. In particular, the heavier
particles carry typically more kinetic energy (even if this is not always the
case). Moreover, the violation of equipartition increases with the mass ratio,
but depends only weakly on the relative densities.

%%%%%%%%%%%%%%%%%%%%%%%%%%%%%%%%%%%%%%%%%%%%%%%%%%%%%%%%%%%%%%%%%%%%%%
\section{Conclusion}
\label{sec:concl}

Because of the dissipative nature of the collisions, granular gases 
are inherently out of equilibrium. They are subject to continuous 
injection and dissipation of energy. An analogy with molecular gases
is usually drawn because of their dilution and incessant collisions.
However, this analogy breaks down as soon as the phenomenology
is studied: numerous effects forbidden by thermodynamics in equilibrium
gases appear (``Maxwell Demon'' experiment, modification of Fourier's
law, non-equipartition of energy) as complex consequences of an
apparent simple ingredient, the inelasticity of collisions.

In this short review, we have arbitrarily chosen to present some aspects
of the rich phenomenology of granular gases, leaving aside many
questions worth of attention. Among these, we can cite the
problem of inelastic collapse \cite{McNamara}, 
the propagation of shock waves \cite{Swinney},
possible phase transitions \cite{Santos}, the formation
of clusters \cite{Falcon2}, the long-range correlations of
hydrodynamic fields \cite{Pre}, thermal
convection \cite{ramirez,wildman,kumaran,viot,meerson1,convPRE}$\ldots$

In conclusion,  in
spite of impressive recent advances in the comprehension
of its phenomenology, the field of granular gases 
still poses serious experimental and theoretical challenges.
In particular, independently of its successes and the technical difficulties,
the hydrodynamic approach is still a controversial issue. 
In view of the lack of scale separation and
neglect of certain correlations, its efficiency seems to go 
(far) beyond what one could have  expected a priori.


\begin{thebibliography}{99}
\bibitem{rque}
The energy is most frequently supplied through a moving boundary
such as a piston, but "bulk" forcings have also been used,
such as the electric field device of \cite{Aranson}.
\end{thebibliography}
\end{document}